\documentclass[mathleft
]{an}
\usepackage{graphicx}

\usepackage{times}
\begin{document}

\Pagespan{1}{}
\Yearpublication{2013}%
\Yearsubmission{2013}%
\Month{07}%
\Volume{999}%
\Issue{88}%

\title{HARPS spectropolarimetry of Herbig Ae/Be stars\thanks
{Based on data obtained from the ESO Science Archive Facility under requests MSCHOELLER 51301, 51324, 36608--36611.}}

\date{Received date / Accepted date}

\author{
{S. Hubrig\inst{1}\fnmsep\thanks{Corresponding author: \email{shubrig@aip.de}}}
\and
I.~Ilyin\inst{1}
\and
M. Sch\"oller\inst{2}
\and
G.~Lo Curto\inst{2}
}

\titlerunning{HARPS spectropolarimetry of Herbig Ae/Be stars}
\authorrunning{S.\ Hubrig et al.}

\institute{
Leibniz-Institut f\"ur Astrophysik Potsdam (AIP), An der Sternwarte 16, 14482 Potsdam, Germany
\and
European Southern Observatory, Karl-Schwarzschild-Str.\ 2, 85748 Garching bei M\"unchen, Germany
}



\abstract{
Our knowledge of the presence and the strength of magnetic fields in intermediate-mass 
pre-main-sequence stars remains very poor. 
We present new magnetic field measurements in six Herbig Ae/Be stars observed with HARPS in 
spectropolarimetric mode.
We downloaded from the European Southern Observatory (ESO) archive the publically available  
HARPS spectra for six Herbig Ae/Be stars.
Wavelength shifts between right- and left-hand side circularly polarised spectra were
interpreted in terms of a longitudinal magnetic field $\left<B_{\rm z}\right>$,
using the moment technique introduced by Mathys.
The application of the moment technique to the HARPS spectra allowed us in addition to study 
the presence of the crossover effect and quadratic magnetic fields.
Our search for longitudinal magnetic fields resulted in first detections of weak magnetic fields in the 
Herbig Ae/Be 
stars HD\,58647 and HD\,98922. Further, we confirm the previous tentative detection of a weak magnetic 
field in HD\,104237 by Donati et al.\ and confirm the previous detection of a magnetic field in the 
Herbig Ae star HD\,190073. Surprisingly, the measured longitudinal magnetic field of HD\,190073,
$\left< B_{\rm z} \right>=91\pm18$\,G at a significance level of 5$\sigma$ is not 
in agreement with the measurement results of Alecian et al. (2013), 
$\left< B_{\rm z} \right>=-10\pm20$\,G, who applied the LSD method to exactly the same data.
No crossover effect was detected for any star in the sample. Only for HD\,98922 the crossover effect was
found to be close to 3$\sigma$ with a  measured value of $-$4228$\pm$1443\,km\,s$^{-1}$\,G.
A quadratic magnetic field of the order of 10\,kG was detected in HD\,98922, 
 and of $\sim$3.5\,kG in HD\,104237.
}

\maketitle

\section{Introduction}
\label{sect:intro}

Herbig Ae/Be stars are pre-main-sequence objects with pronounced emission line 
features and an infrared excess indicative of dust in the circumstellar disks.
Despite the general notion that they do not have 
convectively stable envelopes to support classical T\,Tauri-star-like 
dynamo action, 
in the last years it was demonstrated that a number of Herbig Ae/Be stars have globally 
organised rather strong magnetic fields (e.g., Hubrig et al.\ \cite{Hubrig2004};
Wade et al.\ \cite{Wade2005,Wade2007};
Alecian et al.\ \cite{Alecian2009}; Hubrig et al.\ \cite{Hubrig2009,Hubrig2011a}).

The evidence of large-scale organised magnetic fields detected in Herbig Ae/Be stars is 
reminiscent of the classical Ap/Bp stars, of which the Herbig Ae/Be stars may be precursors.
Magnetic fields in these stars might be fossil of the early star formation
epoch in which the magnetic field of the parental magnetised core was compressed into the innermost 
regions of the accretion disks (e.g.\ Banerjee \& Pudritz \cite{BanerjeePudritz2006}).
Alternatively, Tout \& Pringle (\cite{ToutPringle1995}) proposed a non-solar dynamo that could 
operate in rapidly rotating A-type stars based on rotational shear energy. 

In our recent work based on low-resolution  
polarimetric spectra ($R=2000$) obtained with the multi-mode instrument FORS\,1 at the VLT 
(Hubrig et al.\ \cite{Hubrig2009}), we found a hint that stronger magnetic 
fields appear in very young Herbig Ae/Be 
stars, and that the magnetic fields become very weak at the end of their PMS life, confirming 
the conclusions of Hubrig et al.\ (\cite{Hubrig2000,Hubrig2005,Hubrig2007}) 
that magnetic fields in stars with
 masses less than 3 $M_{\odot}$ are rarely found close to the ZAMS and that kG magnetic fields 
appear in A stars already evolved from the ZAMS.

A series of mean longitudinal magnetic-field measurements for the 
Herbig Ae/Be stars HD\,97048, HD\,150193, and 
HD\,176386 obtained at low resolution with FORS\,2\footnote{The spectropolarimetric 
capabilities of FORS\,1 were moved to FORS\,2 in 2009.} have recently been presented by 
Hubrig et al.\ (\cite{Hubrig2011a}).
This study indicated that dipole models provide a satisfactory fit to the acquired magnetic data.
On the other hand, the work of Adams \& Gregory (\cite{AdamsGregory2011}) shows that high
order field components may even play a dominant role in
the physics of the gas inflow, as the accretion columns approach
the star.

Magnetically controlled behaviour has been found in both the spectroscopic and photometric 
variability of the strongly magnetic Herbig Ae star HD\,101412 (Hubrig et al.\ 
\cite{Hubrig2010,Hubrig2011b}). The small amount of UVES spectra acquired for this 
star have been analysed uncovering variations in equivalent widths, radial velocities, line 
widths, line asymmetries, and mean magnetic field modulus over the rotation period of 42.1\,d.
Clearly, knowledge of the magnetic field structure combined with the 
determination of chemical composition
(e.g., Hubrig et al.\ \cite{Hubrig2010,Hubrig2012};
Cowley et al.\ \cite{Cowley2010}; Folsom et al.\ \cite{Folsom2012}) 
are indispensable to constrain theories on star formation and magnetospheric accretion
in intermediate-mass stars. 

Among the Herbig Ae/Be stars studied 
at low resolution with FORS\,1 and reported as magnetic in previous studies, 
three Herbig Ae/Be stars, HD\,97048, HD\,100546, and HD\,190073, have been observed during the last years 
at single epochs at very high spectral
resolution ($R=115,000$) with the  HARPS spectropolarimeter. They became now publically available in 
the ESO archive. Additionally, for three more Herbig Ae/Be stars, HD\,58647, HD\,98922, and 
HD\,104237, single-epoch high-resolution HARPS spectropolarimetric observations have been acquired
during the last years.

In this study we present our analysis of magnetic fields in this sample of six Herbig Ae/Be stars 
with spectral types between B9 and A4 and compare the measurements with those obtained 
from FORS\,1/2 observations.
For the analysis we use the moment technique developed by 
Mathys (e.g., Mathys \cite{Mathys1991,Mathys1995a,Mathys1995b}), which allows us not only the determination of the 
mean longitudinal magnetic field, but 
also to prove the presence of crossover effect and quadratic magnetic fields.
Conveniently, among the sample stars, one star, HD\,190073, has been 
reported as magnetic  using the so-called Least-Squares Deconvolution (LSD) technique 
by Catala et al.\ (\cite{Catala2007}). Another star,
HD\,104237, was studied with the same technique by Donati et al.\ (\cite{Donati1997}), who achieved a marginal
magnetic detection. These stars are used as a consistency check of our measurements.

\section{Reduction of HARPS spectra and notes on the sample}
\label{sect:descr}

\begin{table*}
\centering
\caption{
The list of the Herbig Ae/Be stars observed with HARPSpol during the last years. 
The spectral types for five targets are taken from SIMBAD. For HD\,104237, the spectral type 
was taken from Mora et al.\ (\cite{Mora2001}).}
\label{tab:objects}
\centering
\begin{tabular}{rccrrrr}
\hline
\hline
\multicolumn{1}{c}{HD} &
\multicolumn{1}{c}{Other} &
\multicolumn{1}{c}{Spectral} &
\multicolumn{1}{c}{$m_V$} &
\multicolumn{1}{c}{MJD} &
\multicolumn{1}{c}{$v\,\sin\,i$} &
\multicolumn{1}{c}{SNR} \\
\multicolumn{1}{c}{number} &
\multicolumn{1}{c}{Identifier} &
\multicolumn{1}{c}{type} &
\multicolumn{1}{c}{} &
\multicolumn{1}{c}{} &
\multicolumn{1}{c}{[km\,s$^{-1}$]} &
\multicolumn{1}{c}{} \\
\hline
58647  & BD$-$13\,2008 & B9IV & 6.85 & 55906.2225 &    118$^1$ & 437 \\
97048  & CD$-$76\,488  & A0   & 8.46 & 55707.9944 &    140$^2$ & 177 \\
98922  & CD$-$52\,4340 & B9V  & 6.76 & 55706.0129 &     50$^3$ & 379 \\
100546 & CD$-$69 893   & B9V  & 6.70 & 55707.0335 &     65$^4$ & 342 \\
104237 & DX\,Cha       & A4   & 6.59 & 55319.2141 &      8$^5$ & 170 \\
190073 & BD\,05\,4393  & A2IV & 7.73 & 55705.4291 & 0--8.6$^6$ & 162 \\
\hline
\end{tabular}
\begin{flushleft}
Notes:\\
Sources for the $v\,\sin\,i$-values are:
$^1$Mora et al.\ (\cite{Mora2001}),
$^2$B\"ohm \& Catala (\cite{BoehmCatala1995}),
$^3$Alecian et al.\ (\cite{Alecian2013a}), 
$^4$Donati et al., (\cite{Donati1997}),
$^5$Cowley et al.\ (\cite{Cowley2013}),
$^6$Catala et al.\ (\cite{Catala2007}).
\end{flushleft}
\end{table*}

The data for all six Herbig Ae/Be stars have been obtained in the polarimetric configuration of HARPS 
(Snik et al.\ \cite{Snik2011}), yielding a spectral 
resolution of 115,000. The targets for which we present spectropolarimetric measurements, together 
with their visual magnitudes,
dates of observations, projected rotational velocities, and the achieved signal-to-noise ratios (SNR)
are presented in Table~\ref{tab:objects}. The HARPS archive spectra cover the wavelength 
range 3780--6913\,\AA{}, with a small gap around 5300\,\AA{}. All spectra were recorded as 
sequences of individual subexposures taken at four different orientations of the quarter-wave 
retarder plate relative to the 
beam splitter of the polarimeter. The reduction was performed using the HARPS data reduction 
software available at the ESO headquarters in Germany. 

\begin{figure*}
\centering
\includegraphics[width=0.35\textwidth]{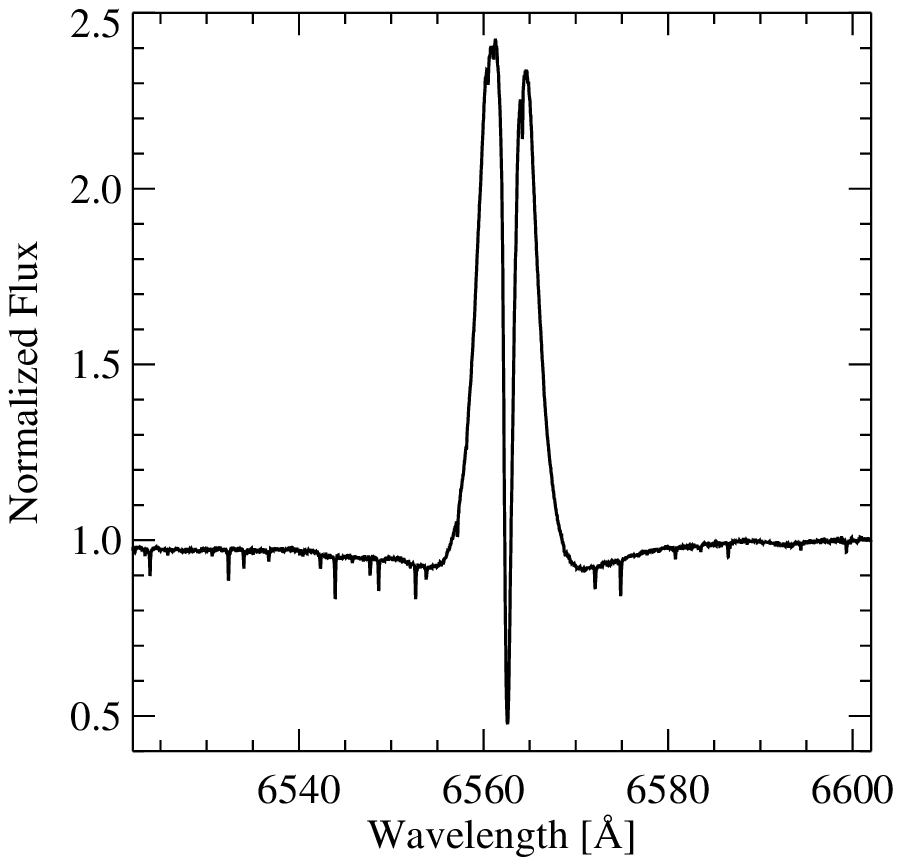}
\includegraphics[width=0.35\textwidth]{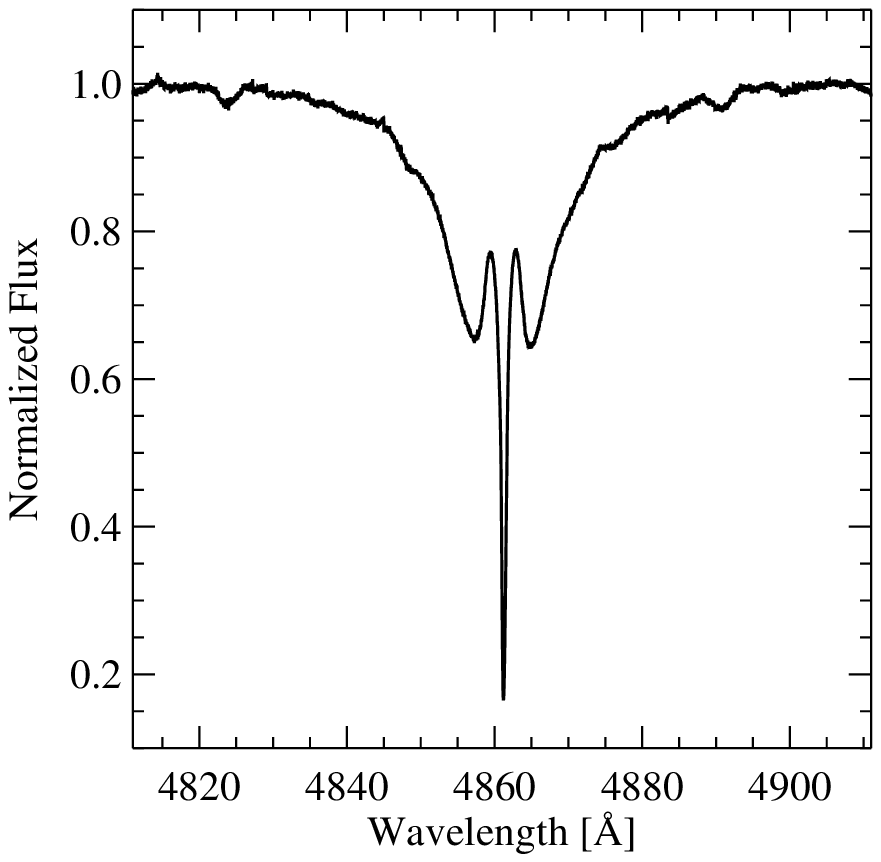}
\caption{Line profiles of H$\alpha$ (left) and H$\beta$ (right) with deep sharp central absorption 
components in the HARPS spectrum of the Herbig Ae/Be star HD\,58647.
}
\label{fig:hydr_58647}
\end{figure*}
To normalise the HARPS spectra to the continuum level, we used the image of the extracted echelle orders packed 
in one FITS file. First, we fit a continuum spline in columns of the image in cross-dispersion direction. 
Each column is fitted in a number of subsequent iterations until it converges to the same upper envelope of 
the continuum level. After each iteration, we analyze the residuals of the fit and make a robust estimation 
of the noise level based upon a statistical test of the symmetric part of the distribution. All pixels whose 
residuals are below the specified sigma clipping level are masked out from the subsequent fit. This way the smooth 
spline function is rejecting all spectral lines below, but leaving the continuum pixels to fit. Once all columns 
are processed, we fit the resulting smoothed curves in the dispersion direction by using the same approach with 
the robust noise estimation from the residuals, but this time rejecting possible outliers above and below the
specified sigma clipping level. As a result, we create a bound surface with continuous first derivatives 
in the columns and rows. We employ a smoothing spline with adaptive optimal regularisation parameters, which 
selects the minimum of the curvature integral of the smoothing spline. As a test for the validity of the 
continuum fit, we check whether the normalised overlapping echelle orders are in good agreement with each other. 
The same is applied to the very broad hydrogen lines, whose winds may span over two or  even three spectral 
orders. The typical mismatch between the red and blue ends of the neighboring orders is well within the statistical 
noise of these orders. The usual procedure to normalise a series of polarimetric observations of the same target 
but with different angles of the retarder, is to create a sum of the individual observations, normalise it to the 
continuum in the way described above, and to use the master normalised image as a template for the individual 
observations: by taking the ratio and fitting a regular spline to it, which then finally defines the 
continuum surface for the individual observation.

HD\,58647 is probably the youngest Herbig Ae/Be star in our sample.
Mendigut\'ia et al.\ (\cite{Mendigutia2012}) assign 
to this star $T_{\rm eff}=10,500$\,K and an age of 0.4\,Myr.
According to the classification scheme of Meeus et al.\ (\cite{Meeus2001}), this 
star with a self-shadowed disk belongs to group~II.
Mendigut\'ia et al.\ (\cite{Mendigutia2011}) did not
detect any spectroscopic variability of 
H$\alpha$, [\ion{O}{i}]~6300, \ion{He}{i}~5876 and \ion{Na}{i}~D lines on three consecutive nights, 
but noted in their work that
variability on timescales longer than three days cannot be excluded. 
The profile of the  H$\alpha$ line from the HARPS spectrum
presented on the left side in Fig.~\ref{fig:hydr_58647} exhibits a somewhat different 
shape compared to that displayed
in the work of Mendigut\'ia et al.\ in Fig.~B.1. The profile is double-peaked with a much deeper sharp 
central absorption 
component. Such a deep sharp central absorption component is also evident in the H$\beta$ profile displayed
on the right side of Fig.~\ref{fig:hydr_58647}. Clearly, additional high-resolution spectroscopic 
observations are needed to solve the issue of variability.


\begin{figure*}
\centering
\includegraphics[width=0.35\textwidth]{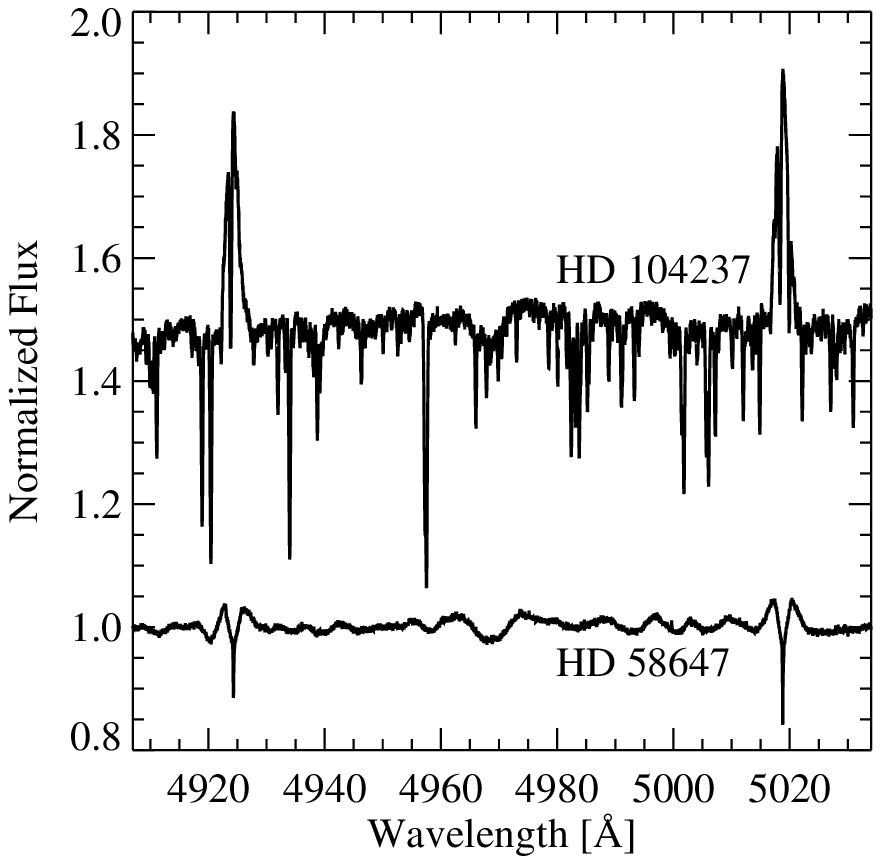}
\includegraphics[width=0.35\textwidth]{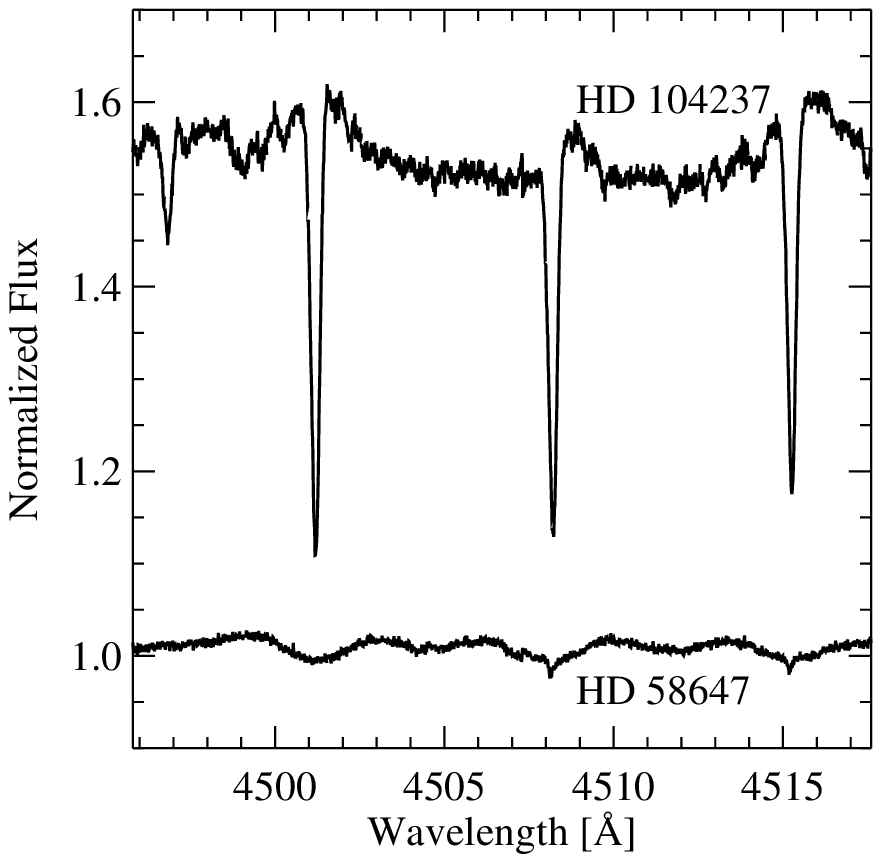}
\caption{Sharp absorption components in the HARPS spectrum of HD\,58647 (lower spectrum).
{\sl Left:} Strong absorption components in the cores of the CS sensitive \ion{Fe}{ii} multiplet 42 
lines at 4923\,\AA{} and 5018\,\AA{}. 
{\sl Right:} Less deep features in the cores of the \ion{Fe}{ii} lines
at 4508\,\AA{} (multiplet 38) and 4515\,\AA{} (multiplet 37).
Note the complete absence
of such a feature in the \ion{Ti}{ii} multiplet 31 line at 4501\,\AA{}. 
To simplify the identification of the sharp features in the cores
of the iron lines, we display in the upper spectra the same spectral region
in the sharp-lined Herbig Ae star HD\,104237.
Please note that the upper spectrum is shifted in vertical direction for clarity.
}
\label{fig:contam_58647}
\end{figure*}

Linear spectropolarimetric data of Mottram et al.\ (\cite{Mottram2007}) reveal that the H$\alpha$ polarisation
behaviour is very different in Herbig Be stars from that of Herbig Ae or T\,Tauri stars. Only the Herbig Be star 
HD\,58647 displays a line-effect at H$\alpha$ and  shows a loop in the QU diagram, similar to Herbig Ae stars within
a magnetospheric accretion scenario. On the other hand, there are hints in the literature that HD\,58647 
is significantly older and could be a Vega-like star (Malfait et al.\ \cite{Malfait1998}) or a 
classical Be star (Manoj et al.\ \cite{Manoj2002}).
The HARPS spectrum shows a fairly rapidly rotating star 
with strong, sharp and deep absorption components  
of \ion{Ca}{ii} H and K centered within the normal, broadened H and K lines, very similar to 
those detected in the star $\beta$\,Pictoris (e.g.\ Slettebak \cite{Slettebak1975}).
A similar feature is visible in the CH(+) 4232\,\AA{} line and in the \ion{Na}{i} D lines. 
Sharp, but less deep absorption components are also detected in numerous 
\ion{Fe}{ii} lines. A few examples are presented in Fig.~\ref{fig:contam_58647}.
Whether  HD\,58647 is a typical young Herbig Be star or belongs to the group of $\beta$\,Pictoris stars
needs further investigation.


\begin{figure*}
\centering
\includegraphics[width=0.35\textwidth]{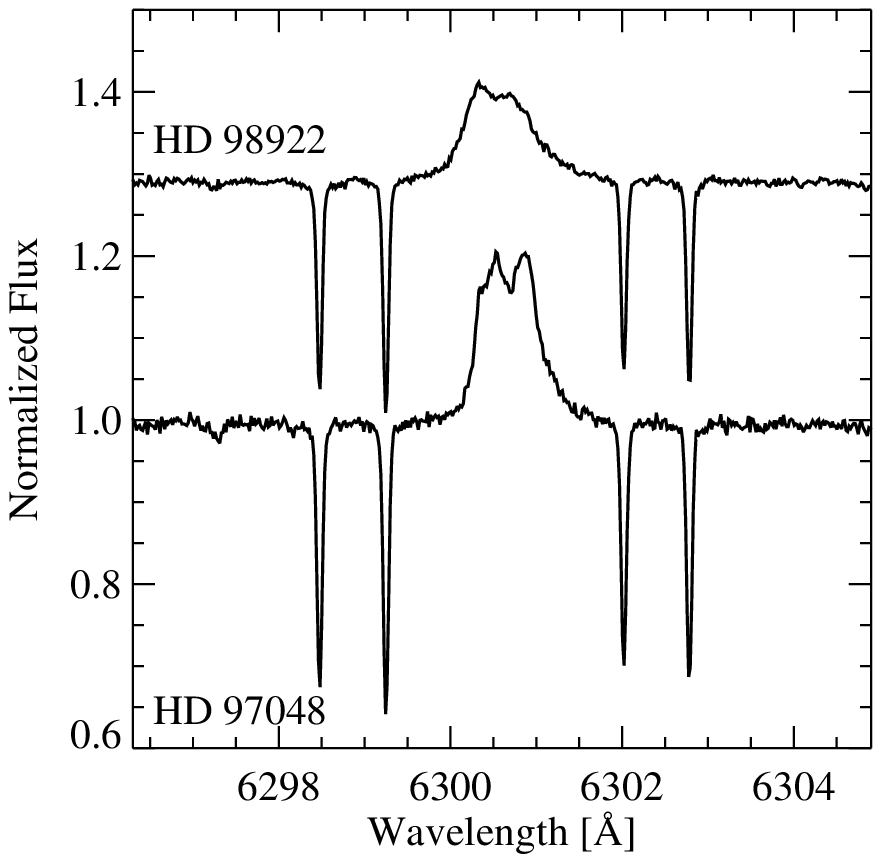}
\includegraphics[width=0.35\textwidth]{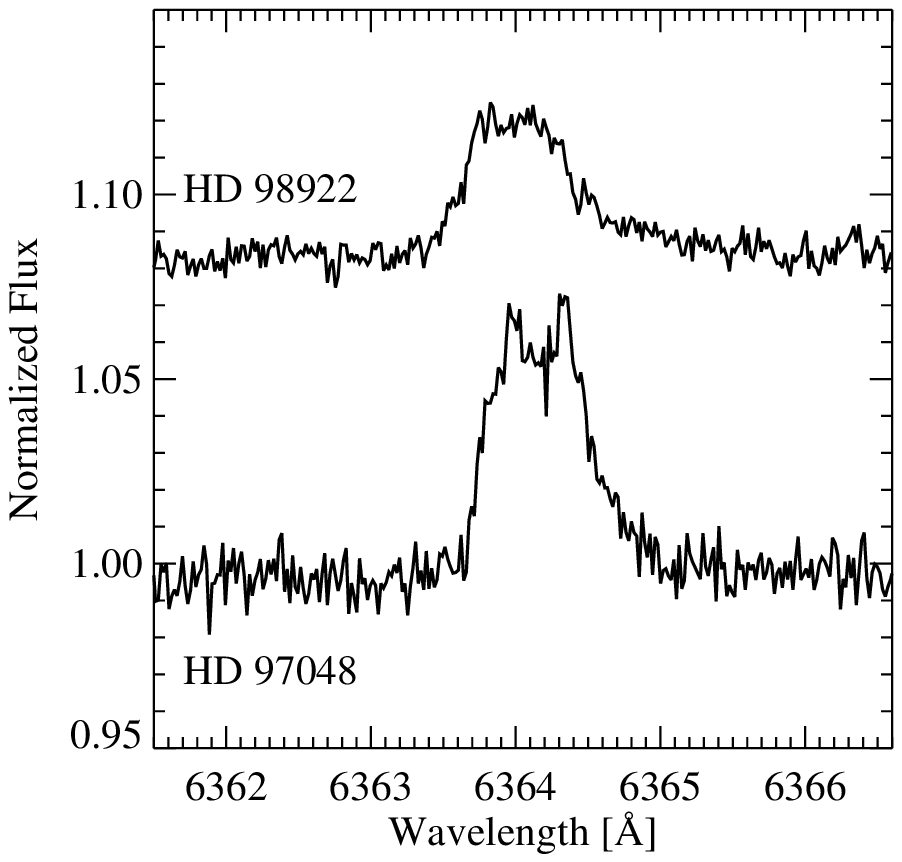}
\caption{
Double-peaked forbidden [\ion{O}{i}] emission line profiles at 6300\,\AA{} and 6363\,\AA{} in the HARPS spectra of 
the Herbig Ae/Be stars HD\,97048 (lower spectra) and HD\,98922 (upper spectra).
Please note that the upper spectra are shifted in vertical direction for clarity.
}
\label{fig:split_o}
\end{figure*}

HD\,97048 is the fastest rotating star in our sample with $v\,\sin\,i=140$\,km\,s$^{-1}$.
This star illuminates the reflection
nebula Ced\,111 and is the brightest member of a cluster of young
stars and the center of ongoing low-mass star formation (Habart et al.\ \cite{Habart2003}).
The disk properties of HD\,97048 have been intensively studied in the recent years.
Acke \& van den Ancker (\cite{AckevandenAncker2006})  resolved extended
[\ion{O}{i}] line emission consistent with a disk with a semi-major axis
oriented at a position angle of 160$^{\circ}$. Similar to their work, also in the HARPS spectrum the [\ion{O}{i}] line emission profiles
are double-peaked. The modeled line profile in the work of Acke \& van den Ancker (\cite{AckevandenAncker2006}) indicates that the double-peaked
lines are caused by the rotation of the disk. In Fig.~\ref{fig:split_o} we present the normalised HARPS spectra
of HD\,97048 and another Herbig Ae/Be star in our sample, HD\,98922,
at the position of forbidden [\ion{O}{i}] emission line profiles. HD\,98922 is the only other Herbig Ae/Be star in our sample
showing double-peaked [\ion{O}{i}] emission lines.

\begin{figure}
\centering
\includegraphics[width=0.35\textwidth]{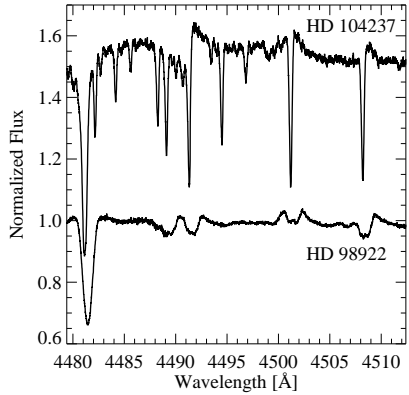}
\caption{Traces of small splitting in a few \ion{Fe}{ii} lines and the \ion{Ti}{ii} $\lambda$4501 line
in the HARPS spectrum of HD\,98922 (lower spectrum).
No splitting is observed in the \ion{Mg}{ii} $\lambda$4481 line. To simplify the identification of 
the spectral lines, we present in the upper part
the HARPS spectrum of the sharp-lined Herbig Ae star  HD\,104237 in the same 
wavelength region.
Please note that the upper spectrum is shifted in vertical direction for clarity.
}
\label{fig:doubling_98922}
\end{figure}

In the HARPS spectrum of HD\,98922 an interesting fact is that the \ion{Fe}{ii} and the \ion{Ti}{ii} lines 
show traces of splitting, best visible in the \ion{Ti}{ii} lines.
Lines belonging to different 
elements exhibit different shapes of their profiles. Such a behaviour suggests that it is possible
that some elements are inhomogeneously
distributed  on the stellar surface. 
A similar behaviour of line profiles was detected in the spectra  of
the strongly magnetic Herbig Ae star HD\,101412 (Hubrig et al.\ \cite{Hubrig2010}), for which the most pronounced 
variability was detected for spectral lines of \ion{He}{i} and the iron peak elements, whereas the spectral 
lines of CNO elements were only slightly variable. In Fig.~\ref{fig:doubling_98922} we present the 
line profiles of \ion{Mg}{ii} $\lambda$4481, a few \ion{Fe}{ii} lines, and the 
\ion{Ti}{ii} $\lambda$4501 line.
The interferometric study of the ${\rm Br}\gamma $ line-emitting region in HD\,98922 by
Kraus et al.\ (\cite{Kraus2008}) 
revealed that this region is compact enough to support the hypothesis that most of 
the ${\rm Br}\gamma $-emission emerges from magnetospheric accretion columns. 

Based on the age estimate of van den Ancker et al.\ (\cite{vandenAncker1998}; $>$10\,Myr), 
HD\,100546 is one of the oldest targets in the sample. 
Near-infrared coronagraphic observations of HD\,100546 detected a disk extending to 515\,AU
(Grady et al.\ \cite{Grady2001}) with an inclination of 49$^{\circ}$.
According to Bouwman et al.\ (\cite{Bouwman2003}), the spectral appearance and dust properties of 
HD\,100546 differ from those of other Herbig Ae/Be stars, showing a large fraction of forsterite and 
an almost identical grain composition as the comet Hale-Bopp. 
Modeling of the SED (Bouwman et al.\ \cite{Bouwman2003}) and imagery with
the Hubble Space Telescope (Grady et al.\ \cite{Grady2005}) indicate the
presence of an inner hole in the dust distribution at 10--13\,AU.
Also the study of the spectral profile of the [\ion{O}{i}] $\lambda$6300 line by Acke \& van den Ancker (\cite{AckevandenAncker2006}) 
suggested that a gap is 
present at 10\,AU in the disk around this star, and such a gap is likely planet-induced. Recent 
high-contrast observations of the circumstellar environment of HD\,100546 by Quanz et al.\ (\cite{Quanz2013})
are interpreted by the presence of a planet in the process of forming.
 
 The Herbig Ae/Be star HD\,104237 was intensively studied during the last years using multi wavelength observations,
in particular due to the possible presence of a magnetic field announced more than 15 years ago by Donati et al.\ (\cite{Donati1997}).
In their work the authors reported on the probable detection of a weak magnetic field of the order of 50\,G.
Imaging observations by Grady et al.\ (\cite{Grady2004}) suggested that the disk of HD\,104237 is seen nearly face-on.
The star is a primary in an SB2 system with an orbital period of 19.86\,d (B\"ohm et al.\ \cite{Boehm2004}). A recent 
study of chemical abundances in both components is presented by Cowley et al.\ (\cite{Cowley2013}).

The absorption and emission spectrum of the Herbig Ae/Be star HD\,190073 was studied in detail by 
Catala et al.\ (\cite{Catala2007}) and more recently by Cowley \& Hubrig (\cite{CowleyHubrig2012}). 
The low projected rotational velocity 
($v\,\sin\,i \le 9$\,km\,s$^{-1}$) may 
indicate either a very slow rotation, or a very small inclination of the rotation axis with 
respect to the line of sight. Recent interferometric observations in the infrared are best 
interpreted in terms 
of a circumstellar disk seen nearly face-on (Eisner et al.\ \cite{Eisner2004}). Also 
IUE observations
indicate a low inclination angle of the rotation axis (Hubrig et al.\ \cite{Hubrig2009}).

Baines et al.\ (\cite{Baines2006}) applied the technique of spectro-astrometry to
study the presence of close companions in a sample of Herbig Ae/Be stars, which also 
includes all six Herbig Ae/Be stars studied in this work.
The method is particularly suited to detect
binary companions of emission-line stars and
can detect companions as much as 6-mag fainter than the primary. The presence of a companion 
was detected for HD\,58647
and HD\,98922, while possible binary detections have been mentioned for HD\,104237 and HD\,190073.
According to the work of B\"ohm et al.\ (\cite{Boehm2004}) and 
Cowley et al.\ (\cite{Cowley2013}), HD\,104237 is a SB2 system where
lines of the secondary are clearly identified in the spectrum.  
No close companions have been reported for HD\,97048 and HD\,100546.

\section{Magnetic field measurements}
\label{sect:meas}

The Stokes~$I$ and $V$ parameters were derived following the ratio method described by 
Donati et al.\ (\cite{Donati1997}), 
ensuring in particular that all spurious signatures are removed at first order. 
Null polarisation spectra (labeled with $null$ in Table~\ref{tab:meas}) have been 
calculated by combining the 
sub-exposures in such a way that the polarisation cancels out, allowing us to verify 
that no spurious signals are present in the data.

\begin{table*}
\centering
\caption{
Line list for the different elements used in the magnetic field measurements.
}
\label{tab:linelist}
\centering
\begin{tabular}{rclrclrcl}
\hline
\hline
\multicolumn{1}{c}{Wavelength} &
\multicolumn{1}{c}{Land\'e} &
\multicolumn{1}{c}{Element} &
\multicolumn{1}{c}{Wavelength} &
\multicolumn{1}{c}{Land\'e} &
\multicolumn{1}{c}{Element} &
\multicolumn{1}{c}{Wavelength} &
\multicolumn{1}{c}{Land\'e} &
\multicolumn{1}{c}{Element} \\
\multicolumn{1}{c}{[\AA{}]} &
\multicolumn{1}{c}{Factor} &
\multicolumn{1}{c}{} &
\multicolumn{1}{c}{[\AA{}]} &
\multicolumn{1}{c}{Factor} &
\multicolumn{1}{c}{} &
\multicolumn{1}{c}{[\AA{}]} &
\multicolumn{1}{c}{Factor} &
\multicolumn{1}{c}{} \\
\hline
4002.5430 & 1.218 & \ion{Fe}{ii} & 4294.0991 & 1.203 & \ion{Ti}{ii} & 4571.9678 & 0.944 & \ion{Ti}{ii} \\
4005.2410 & 1.484 & \ion{Fe}{i}  & 4307.9019 & 1.121 & \ion{Fe}{i}  & 4576.3398 & 1.184 & \ion{Fe}{ii} \\
4012.3850 & 0.716 & \ion{Ti}{ii} & 4383.5439 & 1.149 & \ion{Fe}{i}  & 4588.1992 & 1.059 & \ion{Cr}{ii} \\
4057.4609 & 0.890 & \ion{Fe}{ii} & 4385.3872 & 1.330 & \ion{Fe}{ii} & 4592.0488 & 1.201 & \ion{Cr}{ii} \\
4122.6680 & 1.326 & \ion{Fe}{ii} & 4399.7720 & 1.400 & \ion{Ti}{ii} & 4616.6289 & 0.793 & \ion{Cr}{ii} \\
4132.0581 & 1.510 & \ion{Fe}{i}  & 4404.7500 & 1.129 & \ion{Fe}{i}  & 4618.8032 & 0.914 & \ion{Cr}{ii} \\
4134.6758 & 1.194 & \ion{Fe}{i}  & 4415.1221 & 1.104 & \ion{Fe}{i}  & 4620.5210 & 1.305 & \ion{Fe}{ii} \\
4143.4150 & 1.032 & \ion{Fe}{i}  & 4416.8301 & 0.767 & \ion{Fe}{ii} & 4629.3389 & 1.314 & \ion{Fe}{ii} \\
4145.7808 & 1.173 & \ion{Cr}{ii} & 4464.4502 & 0.493 & \ion{Ti}{ii} & 4634.0698 & 0.508 & \ion{Cr}{ii} \\
4175.6362 & 1.145 & \ion{Fe}{i}  & 4489.1831 & 1.386 & \ion{Fe}{ii} & 4635.3159 & 1.042 & \ion{Fe}{ii} \\
4202.0288 & 1.175 & \ion{Fe}{i}  & 4491.4048 & 0.421 & \ion{Fe}{ii} & 4666.7578 & 1.513 & \ion{Fe}{ii} \\
4242.3638 & 1.200 & \ion{Cr}{ii} & 4508.2881 & 0.503 & \ion{Fe}{ii} & 4731.4531 & 0.655 & \ion{Fe}{ii} \\
4250.1182 & 1.502 & \ion{Fe}{i}  & 4515.3389 & 1.044 & \ion{Fe}{ii} & 4755.7271 & 1.058 & \ion{Mn}{ii} \\
4260.4731 & 1.591 & \ion{Fe}{i}  & 4520.2241 & 1.337 & \ion{Fe}{ii} & 4779.9849 & 1.375 & \ion{Ti}{ii} \\
4271.7588 & 1.236 & \ion{Fe}{i}  & 4522.6338 & 0.921 & \ion{Fe}{ii} & 4812.3369 & 1.499 & \ion{Cr}{ii} \\
4275.5669 & 0.922 & \ion{Cr}{ii} & 4541.5239 & 0.774 & \ion{Fe}{ii} & 4824.1270 & 1.339 & \ion{Cr}{ii} \\
4282.4019 & 1.347 & \ion{Fe}{i}  & 4558.6499 & 1.161 & \ion{Cr}{ii} & 4923.9268 & 1.695 & \ion{Fe}{ii} \\
4284.1880 & 0.521 & \ion{Cr}{ii} & 4563.7612 & 0.985 & \ion{Ti}{ii} & 5018.4399 & 1.933 & \ion{Fe}{ii} \\
\hline
\end{tabular}
\end{table*}


\begin{figure}
\centering
\includegraphics[width=0.45\textwidth]{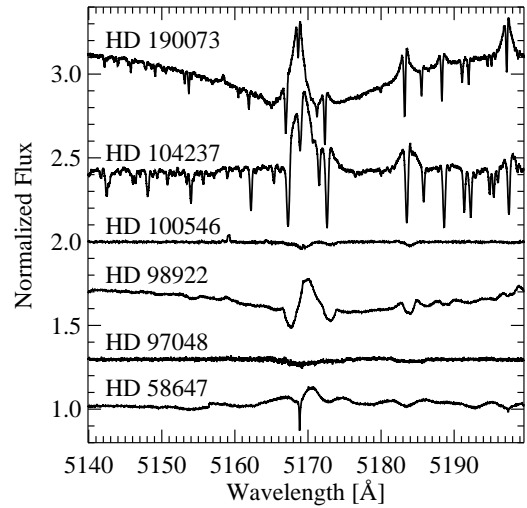}
\caption{The behaviour of the \ion{Fe}{ii} multiplet 42 line at $\lambda$5169 in all studied stars.
From bottom to the top we present the HARPS spectra of HD\,58647, HD\,97048, HD\,98922, HD\,100546, 
HD\,104237, and HD\,190073.
Please note that the spectra are shifted in vertical direction for clarity.
}
\label{fig:herbig_all}
\end{figure}

Wavelength shifts between right- and left-hand side circularly polarised spectra were
interpreted in terms of a longitudinal magnetic field $\left<B_{\rm z}\right>$,
using the moment technique described by Mathys (\cite{Mathys1994}).
All the lines we tried to employ in the diagnosis
of the magnetic fields on the surface of our target stars are presented in Table~\ref{tab:linelist},
together with their Land\'e factors. 
The wavelengths and
the Land\'e factors were 
taken from Kurucz's  list of atomic data (Kurucz \cite{Kurucz1989}). 
We note that the actual list of lines measured in the
spectrum of each star
can differ from one observation to the next due to the different quality 
of the spectra, contamination by the CS material, and the projected rotational 
velocity where blend-free spectral lines are difficult to identify in fast rotating targets.
All stars have very different line widths, with the sharpest lines in HD\,190073 and HD\,104237.
The other four stars have $v\,\sin\,i$ values $\ge50$\,km\,s$^{-1}$.
The contamination by the circumstellar (CS) material is different in different stars.
The CS environment of the Herbig Ae/Be stars typically consist 
of a combination of disk, wind, accretion, and jets.
As an example of the different impact of the CS material on the line profiles in our sample stars, 
we present in Fig.~\ref{fig:herbig_all} the spectral region with the profiles of 
the \ion{Fe}{ii} $\lambda$5169 lines  of multiplet 42.  

\begin{table}
\centering
\caption{
Measurements of the mean longitudinal magnetic field in our sample stars.
All quoted errors are 1$\sigma$ uncertainties.
}
\label{tab:meas}
\centering
\begin{tabular}{lr@{$\pm$}lr@{$\pm$}l}
\hline
\hline
\multicolumn{1}{c}{Object \rule[-1.6ex]{0pt}{4.2ex}} &
\multicolumn{2}{c}{$\left< B_{\rm z} \right>$} &
\multicolumn{2}{c}{$\left< B_{\rm z} \right>_{\rm null}$} \\
\multicolumn{1}{c}{} &
\multicolumn{2}{c}{[G]} &
\multicolumn{2}{c}{[G]} \\
\hline
HD\,58647   &    218 & 69 &    54 & 71 \\
HD\,98922   & $-$131 & 34 & $-$42 & 36 \\
HD\,104237  &     63 & 15 &  $-$5 & 17 \\
HD\,190073  &     91 & 18 &    21 & 18  \\
\hline
\end{tabular}
\end{table}

The measurements of the mean longitudinal magnetic field $\left<B_{\rm z}\right>$ together with 
the mean longitudinal 
magnetic field determined from null spectra $\left< B_{\rm z} \right>_{\rm null}$ using HARPS 
spectropolarimetric observations are listed in Table~\ref{tab:meas}. 
The presence of a weak positive longitudinal magnetic field $\left< B_{\rm z} \right>=218\pm69$\,G 
at a significance level of 3.2$\sigma$ in HD\,58647 is reported here for the first time. 
Due to the presence of sharp narrow components 
in a number of spectral lines, only  the lines without visually  noticeable contribution of 
circumstellar material have been selected 
for magnetic field measurements.

\begin{figure}
\centering
\includegraphics[width=0.35\textwidth]{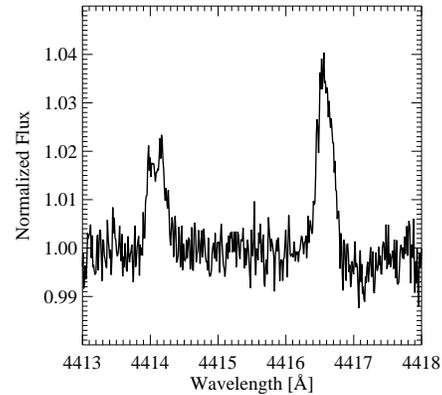}
\caption{Forbidden [\ion{Fe}{ii}] emission lines of multiplets 6F and 7F in the wavelength region
4413--4417\,\AA{} in the HARPS spectrum of HD\,100546. Note that the double-peaked profile at $\lambda$4414 consists of two emission lines,
$\lambda$4413.79 (multiplet 7F) and $\lambda$4414.45 (multiplet 6F).
}
\label{fig:emis_hd100_Fe}
\end{figure}

\begin{figure}
\centering
\includegraphics[width=0.35\textwidth]{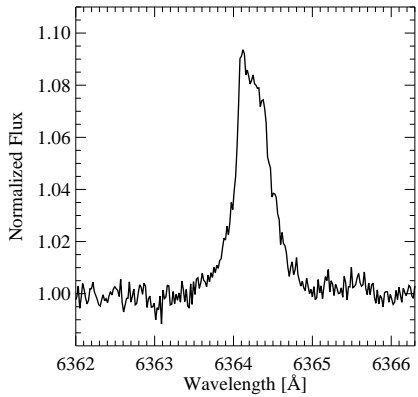}
\caption{Asymmetric line profile of the forbidden [\ion{O}{I}] emission line at 6363\,\AA{} in the HARPS 
spectrum of HD\,100546.
}
\label{fig:emis_hd100_O}
\end{figure}

We were not able to determine a longitudinal magnetic field 
in the HARPS spectra of the Herbig Ae/Be stars 
HD\,97048 and  HD\,100546. The HARPS spectrum of HD\,97048 retrieved from the ESO archive
was obtained at a rather low SNR. Since HD\,97048 is the  fastest rotating target in our sample,
a high SNR, up to a few hundreds, is indispensable to detect a magnetic field.
As for HD\,100546, we experienced the same surprise  as  Donati et al.\ (\cite{Donati1997}) 
finding only very few photospheric features in the spectrum, but a number of forbidden
emission [\ion{Fe}{ii}] lines.
In addition to the forbidden emission [\ion{Fe}{ii}] lines belonging to multiplets 19F and 20F
identified by Donati et al.\ (\cite{Donati1997}), in our HARPS spectrum we have detected also forbidden
emission [\ion{Fe}{ii}] lines belonging to multiplets 6F and 7F.
Some of the detected emission lines appear very symmetric, although there is a number of line profiles 
of [\ion{Fe}{ii}] in the 
HARPS spectrum showing asymmetric line profiles. 
Also the forbidden emission [\ion{O}{i}] lines show slightly asymmetric profiles.
In Fig.~\ref{fig:emis_hd100_Fe} we present the spectral region around the forbidden
emission [\ion{Fe}{ii}] lines belonging to multiplets 6F and 7F. An example of an asymmetric 
profile of the forbidden 
emission [\ion{O}{i}] line at $\lambda$6393 is displayed in Fig.~\ref{fig:emis_hd100_O}. No 
magnetic field was detected in HD\,100546 by Donati et al.\ (\cite{Donati1997}) using the LSD 
technique. Our magnetic field measurements 
using just a few photospheric lines cannot be considered reliable due to the large 
measurement inaccuracy and are not presented in Table~\ref{tab:meas}.

On the other hand, Hubrig et al.\ (\cite{Hubrig2009,Hubrig2011a}) reported the detection of 
weak longitudinal magnetic fields in both Herbig Ae/Be stars HD\,97048 and HD\,100546 using 
low-resolution spectra obtained in spectropolarimetric mode with FORS\,1/2 mounted 
on the VLT.  
Using multi-epoch polarimetric spectra for HD\,97048, the authors detected a variation
of the longitudinal magnetic field in the range from $-186$\,G to 164\,G and  
suggest a magnetic/rotation period of 0.693\,d. The longitudinal magnetic field of HD\,100546
appears to be much weaker, only of the order of 90\,G. Clearly, additional high-quality 
spectropolarimetric observations are needed to confirm or disprove the presence of a magnetic
field in these stars.

Also for the first time we present here the detection of a longitudinal magnetic field
$\left< B_{\rm z} \right>=-131\pm34$\,G at a significance level of 3.9$\sigma$ in the 
moderately rotating Herbig Ae/Be star HD\,98922.
This star was previously observed only once by Wade et al.\ (\cite{Wade2007}) who have not
detected the presence of a magnetic field using low-resolution FORS\,1 spectra. 

The best accuracy in the magnetic field 
determination using HARPS spectra is achieved in both sharp-lined stars HD\,104237 and HD\,190073. 
Donati et al.\ (\cite{Donati1997}) detected a marginal Zeeman signature in the LSD Stokes $V$
profile of HD\,104237 in the spectra obtained with a visitor polarimeter 
coupled to the UCL Echelle Spectrograph at AAT. Our measurement, $\left<B_{\rm z}\right>=63\pm15$\,G,
is the first confirmation of the presence of a magnetic field in this star.

The first measurement of a longitudinal magnetic field in HD\,190073 was published by 
Hubrig et al.\ (\cite{Hubrig2006})
indicating the presence of a longitudinal magnetic field $\left<B_{\rm z}\right> =84\pm30$\,G measured 
on FORS\,1 low-resolution spectra at 2.8$\sigma$ level.
Later on this star was studied by Catala et al.\ (\cite{Catala2007}) using ESPaDOnS observations, who 
confirmed the presence of a weak magnetic field,  $\left<B_{\rm z}\right> =74\pm10$\,G, at a higher
significance level. 
A few years later a longitudinal magnetic field $\left<B_{\rm z}\right> =104\pm19$\,G  was 
reported by Hubrig et al.\ (\cite{Hubrig2009}) using FORS\,1 measurements. The measurement of 
the longitudinal magnetic field using the available archival HARPS observations from May 2011, 
$\left<B_{\rm z}\right> =91\pm18$\,G presented in this work  
fully confirms the presence of a rather stable weak field. 

Surprisingly, new observations of this star during 
July 2011 and October 2012 by Alecian et al.\ (\cite{Alecian2013b}) detected variations of the 
Zeeman signature in the LSD spectra on timescales of days to weeks.
Among the Zeeman signatures displayed in their Fig.~1, the LSD Zeeman feature extracted from the 
same HARPS spectrum as in our 
analysis is also presented, but the strength of the measured longitudinal magnetic field is only
$\left<B_{\rm z}\right> =-10\pm20$\,G.
The authors suggest that the
detected variations of Zeeman signatures are the result of the interaction between the fossil 
field and the ignition 
of a dynamo field generated in the newly-born convective core. As our measurements completely
contradict those presented by Alecian et al., careful 
spectropolarimetric monitoring over the next years is important to confirm the reported variability
of the magnetic field.

No crossover effect was detected for any star in the sample. 
The crossover effect is measured by the second-order moment about the centre of the profiles of 
spectral lines recorded in the Stokes parameter $V$ (Mathys \cite{Mathys1995a}).
%
%
Only for HD\,98922 the crossover effect was found to be
close to 3$\sigma$ with a measured value of $-$4228$\pm$1443\,km\,s$^{-1}$\,G. 

A quadratic magnetic field of the order
of 10\,kG ($\left<B_{\rm q}\right>=10,246\pm3246$\,G) was detected in HD\,98922, 
 and of $\sim$3.6\,kG ($\left<B_{\rm q}\right>=3562\pm1164$\,G) in HD\,104237.
A quadratic magnetic field close to the 3$\sigma$ level, $\left<B_{\rm q}\right>=3169\pm1132$\,G
was measured for HD\,190073.
The mean quadratic 
magnetic field,
\begin{displaymath}
\langle B_q\rangle= (\langle B^2\rangle + \langle B_z^2\rangle)^{1/2},
\end{displaymath}
is derived through the application of the moment technique, 
as described e.g.\ by Mathys \& Hubrig (\cite{MathysHubrig1995}). Here, $\langle B^2\rangle$ is 
the mean square magnetic field modulus, i.e.\ the average over the stellar 
disc of the square of the modulus of the magnetic field vector, weighted by 
the local emergent line intensity, while $\langle B_z^2\rangle$ is 
the mean square longitudinal magnetic field, i.e.\ the average over the stellar 
disc of the square of the line-of-sight component of the magnetic 
vector, weighted by the local emergent line intensity. 
Importantly, contrary to the 
mean longitudinal magnetic field, the mean quadratic magnetic field provides a measurement of the field strength 
that is fairly
insensitive to its structure.

\section{Discussion}

With this work we increase the number of Herbig Ae/Be stars showing the presence of 
magnetic fields. Longitudinal magnetic fields
have  been detected in about a dozen Herbig Ae/Be stars 
(e.g., Hubrig et al.\ \cite{Hubrig2004,Hubrig2006,Hubrig2007,Hubrig2009,Hubrig2011a,Hubrig2011b};
Wade et al.\ \cite{Wade2007}; Catala et al.\ \cite{Catala2007}; Alecian et al.\ \cite{Alecian2009}).
For the majority of these stars rather small magnetic fields were measured, of the order of only 
100\,G or less. 
Presently, the Herbig Ae star HD\,101412 possesses the strongest 
longitudinal magnetic field ever measured in any Herbig Ae star, 
with a surface magnetic field $\left<B\right>$ up to 3.5\,kG. 
HD\,101412 is also the only Herbig Ae/Be star for which the rotational
Doppler effect was found to be small in comparison to the magnetic splitting 
and  several spectral lines
observed in unpolarised light at high dispersion are resolved into 
magnetically split components (Hubrig et al.\ \cite{Hubrig2010,Hubrig2011b}).
  
To understand the magnetospheres of Herbig Ae/Be stars
and their interaction with the circumstellar environment presenting
a combination of disk, wind, accretion, and jets, the
knowledge of the magnetic field strength and topology is indispensable.
Progress in understanding
the disk-magnetosphere interaction can, however,
only come from studying a sufficient number of targets in
detail to look for patterns encompassing this type of pre-main sequence stars.

In the presented high-resolution HARPS spectra, the contamination of the spectra by the CS material is 
clearly visible, and it is of utmost importance to study the disk-magnetosphere interaction in 
detail by modeling the Stokes~$V$ profiles not only in photospheric lines, but also 
in various wind and accretion diagnostic lines. The Doppler-shifted spectropolarimetric contributions
from photospheric and circumstellar environmental diagnostic lines should be investigated 
using high-resolution polarimetric observations over the stellar rotation period. 
These studies will enable the application of
the technique of Doppler Zeeman tomography to determine the correspondence between the 
magnetic field structure and the radial density and temperature profiles and to construct for the first
time a realistic model of magnetospheres 
of Herbig Ae/Be stars that takes into account the full complexity of the circumstellar environment, 
including the observed outflows and jets.

{
\acknowledgements
This research has made use of the SIMBAD database,
operated at CDS, Strasbourg, France.
}


\label{lastpage}

\end{document}